\documentclass[fleqn,usenatbib]{mnras}

\usepackage{newtxtext}

\usepackage[T1]{fontenc}

\DeclareRobustCommand{\VAN}[3]{#2}
\let\VANthebibliography\thebibliography
\def\thebibliography{\DeclareRobustCommand{\VAN}[3]{##3}\VANthebibliography}

\usepackage{graphicx}	
\usepackage{newtxmath}

\usepackage{amsmath}	
\usepackage{amssymb}	
\usepackage{tabulary}
\usepackage{tablefootnote}
\usepackage{float}

\title[The Extended Schmidt Law of  \ion{H}{I}-rich UDGs]{The  \ion{H}{I}-rich Ultra-diffuse Galaxies follow the Extended Schmidt Law}

\author[Sai Zhai et al.]{
Sai Zhai,$^{1,2}$
Yong Shi,$^{1,2}$ \thanks{E-mail: yong@nju.edu.cn}
Zhi-Yu Zhang,$^{1,2}$
Jun-Zhi Wang,$^{3,4}$
Yu Gao, $^{5,6}$
Qiusheng Gu, $^{1,2}$
\newauthor
Tao Wang, $^{1,2}$
Kaiyi Du, $^{1,2}$
Xiaoling Yu, $^{7,1,2}$
and Xin Li $^{1,2}$
\\
$^{1}$School of Astronomy and Space Science, Nanjing University, Nanjing 210093, China\\
$^{2}$Key Laboratory of Modern Astronomy and Astrophysics (Nanjing University), Ministry of Education, Nanjing 210093, China\\
$^{3}$Shanghai Astronomical Observatory, Chinese Academy of Sciences, 80 Nandan Road, Shanghai 200030 PR China \\
$^{4}$School of Physical Science and Technology, Guangxi University, Nanning 530004, PR China \\
$^{5}$Department of Astronomy, Xiamen University, Xiamen, Fujian 361005, People’s Republic of China \\
$^{6}$Purple Mountain Observatory, Chinese Academy of Sciences, No.10 Yuanhua Road, Qixia District, Nanjing 210023, People’s Republic of China\\
$^{7}$College of Physics and Electronic Engineering, Qujing Normal University, Qujing 655011, P.R. China
}

\date{Accepted XXX. Received YYY; in original form ZZZ}

\pubyear{2015}

\begin{document}
\label{firstpage}
\pagerange{\pageref{firstpage}--\pageref{lastpage}}
\maketitle

\begin{abstract}
The \ion{H}{I}-rich ultra-diffuse galaxies (HUDGs) offer a unique case for studies of star formation laws (SFLs) as they host low star formation efficiency (SFE) and low-metallicity environments where gas is predominantly atomic. We collect a sample of six HUDGs in the field and investigate their location in the extended Schmidt law($\Sigma_{\text {SFR }} \propto \left(\Sigma_{\text{star}}^{0.5} \Sigma_{\text{gas}}\right)^{1.09}$).
They are consistent with this relationship well (with deviations of only 1.1 sigma). Furthermore, we find that HUDGs follow the tight correlation between the hydrostatic pressure in the galaxy mid-plane and the quantity on the x-axis ($\rm log(\Sigma_{star}^{0.5}\Sigma_{gas})$) of the extended Schmidt law. This result indicates that these HUDGs can be self-regulated systems that reach the dynamical and thermal equilibrium. In this framework, the stellar gravity compresses the disk vertically and counteracts the gas pressure in the galaxy mid-plane to regulate the star formation as suggested by some theoretical models.
\end{abstract}




\section{Introduction}

The availability of deep observations, such as Dragonfly telephoto array \citep{2014PASP..126...55A}, has led to increasing interest in ultra-diffuse galaxies (UDGs).
The  \ion{H}{I}-rich ultra-diffuse galaxies (HUDGs) are a sub-group of UDGs. In addition to some typical characteristics of UDGs, such as low surface brightness ($\mu_{g, 0} \gtrsim 24 \ \mathrm{mag} \operatorname{arcsec}^{-2}$), and large optical half light radii ($R_{e} \gtrsim 1.5 \ \mathrm{kpc}$, \cite{2018ApJ...856L..31T}), they are rich in  \ion{H}{I} gas \citep{2017ApJ...842..133L,2020MNRAS.499L..26W}. Although previous studies have found them to be field star forming galaxies \citep{2017A&A...601L..10P,2017ApJ...842..133L, 2021ApJ...909...19G, 2021ApJ...909...20S}, their molecular gas fractions and star formation efficiencies ($\rm SFEs=SFR/M_{ \ion{H}{I}}$) are low \citep{2017ApJ...842..133L,2020MNRAS.499L..26W}. However, the origin of their low SFEs is poorly understood.

Several mechanisms have been proposed to explain the low SFEs of ultra-diffuse galaxies (UDGs).  \cite{2015MNRAS.452..937Y} proposed that they might be failed L* galaxies with suppressed star formation. In this situation, galaxies should fall into dense regions before quenching, such as clusters. However, most of HUDGs are found as isolated galaxies \citep{2017ApJ...842..133L}. Also, if the star formation is suppressed by environments, they should be gas-poor, which is in contrast to the situation that HUDGs are \ion{H}{I}-rich. \cite{2016MNRAS.459L..51A} mentioned that high spin dark matter halos will prevent the collapse of baryons, which also hinders star formation.  \cite{2017MNRAS.466L...1D} pointed out that they could be normal dwarfs with strong stellar feedback which remove the cold gas and thus suppress the star forming processes. This is also in contrast to the current properties of HUDGs. It has been suggested that their extremely low SFE is driven by the low stellar mass surface density, as suggested by the extended Schmidt (ES) law  \citep{2011ApJ...733...87S,2018ApJ...853..149S}.

In this work, we aim to extend the validity of the latter possibility by using both the global and local ES law. Here, global measurements are for unresolved galaxies where quantities are averaged over the entire star-forming disc, and local measurements are for resolved galaxies where measurements can be obtained from azimuthally-averaged radial profiles or pixel-by-pixel.  The most notable distinction between the ES law and other star formation laws (SFL) is the explicit inclusion of the stellar mass contribution to the SFR, expressed as $\Sigma_{\text {SFR }} \propto \left(\Sigma_{\text{star}}^{0.5} \Sigma_{\text{gas}}\right)^{1.09}$. Specifically, the Kennicutt-Schmidt (KS) law demonstrates the primary contribution of gas to the SFR, given by $
\rho_{\mathrm{SFR}}=A \rho_{\mathrm{gas}}^{N}
$ \citep{1959ApJ...129..243S,1989ApJ...344..685K,1998ApJ...498..541K, 2019A&A...632A.127B, 2020A&A...644A.125B}. Another SFL, known as the Silk-Elmegreen (SE) elation 
 establishes a relationship between the SFE and orbital dynamical time ($\tau_{\mathrm{dyn}}$): $\mathrm{SFE} \propto \frac{1}{\tau_{\mathrm{dyn}}}$  \citep{1997ApJ...481..703S, 1997ApJ...486..944E}.
 However, metal-poor galaxies significantly deviate from the KS law and SE relation \citep{2011ApJ...733...87S}. The explict inclusion of stellar component reduces the deviation of metal-poor galaxies from the best-fitting of ES law \citep{2011ApJ...733...87S,2018ApJ...853..149S}. It has been found that the ES law holds across different environments, from star-forming environments, the outer regions of dwarf galaxies to local spiral galaxies and high-redshift objects \citep{2018ApJ...853..149S,2018ApJS..235...18L, 2023MNRAS.518.4024D}, thereby providing a unified explanation to the SFEs. \cite{2021ApJ...909...20S} found that the HUDG AGC 242019 closely follows the extended Schmidt law. Here we expand the sample to six HDUGs to explore whether they follow the extended Schmidt law aiming to understand whether their low SFE is regulated by stellar mass through mid-plane pressures.  We present the sample selection and data in Section 2, the result and discussion in Section 3, and the conclusion in Section 4. We adopt a $\Lambda \rm CDM$ cosmology with $\Omega_{\rm m}$=0.27, $\Omega_{\Lambda}$=0.73, and $H_{0}=73 \rm km \ s^{-1} \ Mpc^{-1}$ \citep{2016A&A...594A..13P}.

\begin{figure*}
    \centering
    \includegraphics[width=\textwidth]{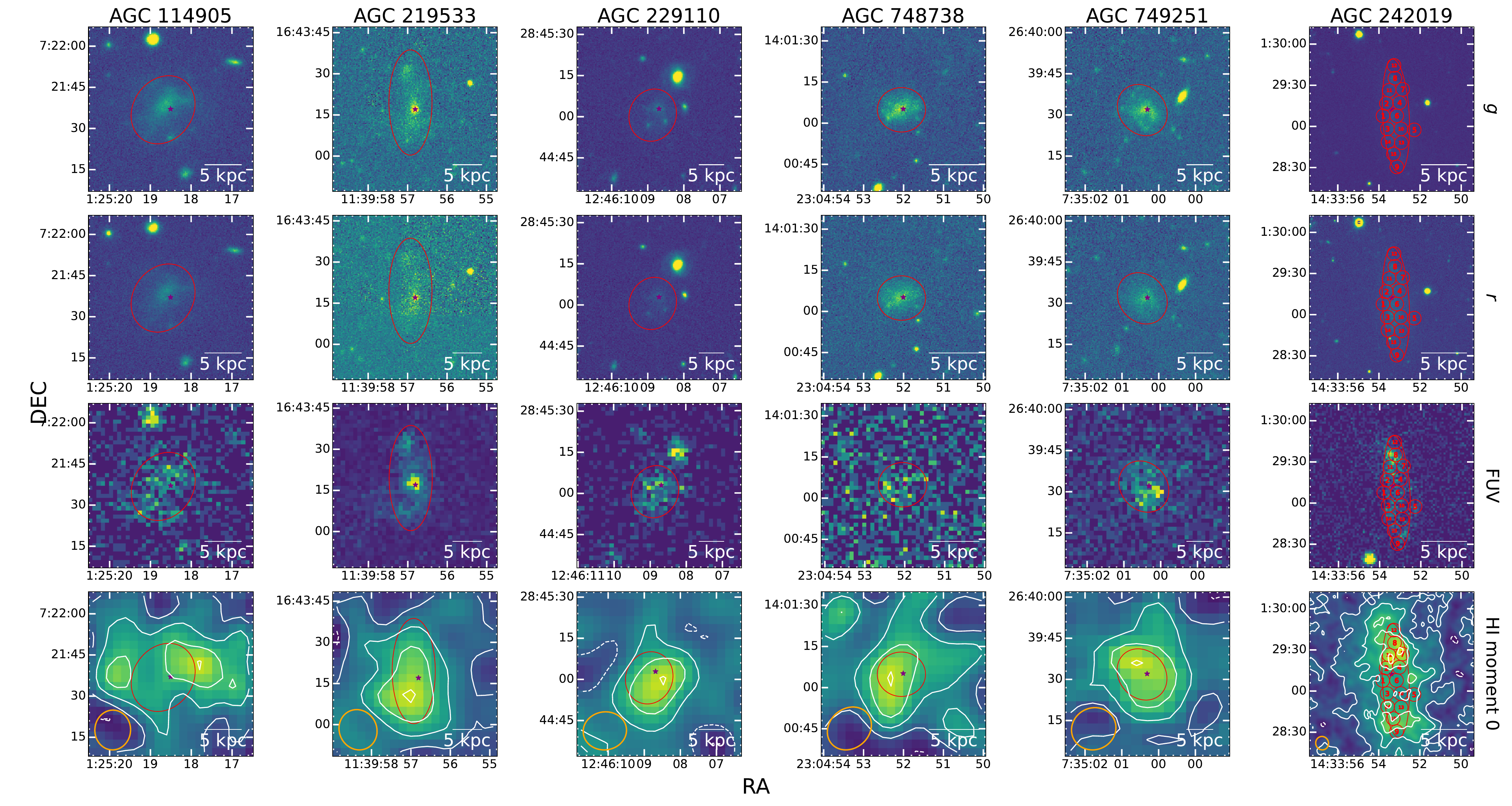}
    \caption{The multi-band images of the six  \ion{H}{I}-rich UDGs. From top to bottom, they are the $g$ band, $r$ band, FUV band, and zeroth-moment map of  \ion{H}{I} observations images, respectively. The x-axis and y-axis show the R.A. and DEC. of each galaxy. The center of galaxies from \citet{2021ApJ...909...19G} and \citet{2021ApJ...909...20S} are the optical emission center ($g,r$) and dynamical center (based on the \ion{H}{I} kinematics), respectively. The major tick spacing represents $15''$ for the first five rows and $30''$ for the rest. The red and orange ellipses represent the aperture sizes and the \ion{H}{I} beam sizes, respectively. The small red ellipse in AGC 242019 image are labeled corresponding to the sequence in Table \ref{tab:properties_resolved}. The white contours are the HI zeroth-moment emission. A scale bar of 5 kpc is shown in the right corner of each panel.}
    \label{fig:multiband}
\end{figure*}

\begin{table*}
\renewcommand\arraystretch{1.5}
\centering
\caption{Global properties of HUDGs}
\begin{tabular}{ccccccccccc}
\hline
\hline
\begin{tabular}[c]{@{}c@{}}AGC ID \\ \end{tabular} & \begin{tabular}[c]{@{}c@{}}RA \\ (J2000) \end{tabular} & \begin{tabular}[c]{@{}c@{}} DEC \\
(J2000) \end{tabular} & \begin{tabular}[c]{@{}c@{}} Dist. \\  (Mpc)
 \end{tabular}  & \begin{tabular}[c]{@{}c@{}}  Redshift \\  
 \end{tabular}  & \begin{tabular}[c]{@{}c@{}}i \\ ($^{\circ}$) \end{tabular} & \begin{tabular}[c]{@{}c@{}} VLA  beam size\\ ('' $\times$ ")\end{tabular}   & \begin{tabular}[c]{@{}c@{}} $\Sigma_{\text{SFR}}$ \\ ($M_{\odot}/\rm{yr}/\rm{kpc^{2}}$) \end{tabular}  & \begin{tabular}[c]{@{}c@{}}$\Sigma_{\text{star}}$\\ ($M_{\odot}/\rm{pc^{2}}$) \end{tabular}  & \begin{tabular}[c]{@{}c@{}}$\Sigma_{\text{atomic}}$\\ ($M_{\odot}/\rm{pc^{2}}$)\end{tabular} & \begin{tabular}[c]{@{}c@{}}References \\ 
\end{tabular} \\ \hline
229110 & 191.5362 & 28.7508 & 112 & 0.46& 28.4 & 15.8 $\times$ 13.8 &
 $2.13^{+0.52}_{-0.27} \times 10^{-4}$ & $1.20^{+0.29}_{-0.15}$ & $4.84^{+1.19}_{-0.60}$ & 1 \\ 
 114905 & 21.3271 & 7.3603 & 76  &0.33 & 33.9 & 14.5 $\times$13.00 &
 $1.30^{+0.38}_{-0.22} \times 10^{-4}$ & $1.57^{+0.46}_{-0.27}$ & $4.23^{+1.23}_{-0.72}$ & 1 \\ 
 219533 & 174.9867 & 16.7214 & 96  & 0.40 & 62.0 & 14.9 $\times$ 13.6 &
 $8.85^{+6.22}_{-5.15}\times 10^{-5}$ & $3.19^{+2.24}_{-1.86}\times 10^{-1}$ & $3.23^{+2.27}_{-1.88}$ & 1 \\ 
748738 & 346.2167 & 14.0181 & 56 & 0.24 & 23.1 & 17.1 $\times$ 14.5 &  $7.69^{+1.58}_{-0.66}\times 10^{-5}$ & $2.91^{+0.60}_{-0.25}$ & $4.02^{+0.83}_{-0.34}$ & 1 \\ 
749251 & 113.7513 & 26.6589 & 106 & 0.43 & 31.8 & 16.4 $\times$15.2  &  $1.37^{+0.37}_{-0.21} \times 10^{-4}$ & $2.21^{+0.60}_{-0.33}$ & $3.84^{+1.05}_{-0.58}$ & 1 \\ 
242019 & 218.4724 & 1.4868 & 30.8 & 0.13 & 73.1 & 9.85 $\times$ 9.33 & $5.39^{+0.86}_{-0.06}\times 10^{-5}$&$5.80^{+0.92}_{-0.07}\times 10^{-1}$&$3.03^{+0.48}_{-0.03}$& 2 \\

  \hline
\\
\end{tabular}
\begin{flushleft}
\textit{Notes}: The distance is calculated with the heliocentric velocity \citep{2011AJ....142..170H,2021ApJ...909...19G,2021ApJ...909...20S}. The inclination of AGC 242019 is determined using the tilted-ring modeling tool as described in \cite{2021ApJ...909...20S} and \cite{2015MNRAS.451.3021D}. For other HUDGs, the inclination angles are obtained by visually determining the ratios of the minor axis and the major axis , based on the contour of the $g$ band image using the DS9 \citep{2003ASPC..295..489J}. $\rm \Sigma_{gas}=\Sigma_{atomic} = 1.36 \times \Sigma_{\ion{H}{I}}$ due to the limited quantity of molecular gas in HUDGs \citep{2020MNRAS.499L..26W}.
The errors come from the systematic errors of the parameters and those from the measurement of the inclination angles.

References: 1. \cite{2021ApJ...909...19G}; 2. \cite{2021ApJ...909...20S}.

\end{flushleft}
\label{tab:properties}
\end{table*}

\begin{table}
\renewcommand\arraystretch{1.5}
\centering
\caption{Spatially-resolved properties of 15 regions from AGC 242019}
\begin{tabular}{ccccccccccc}
\hline
\hline
 \begin{tabular}[c]{@{}c@{}}RA \\ (J2000) \end{tabular} & \begin{tabular}[c]{@{}c@{}} DEC \\
(J2000) \end{tabular} & \begin{tabular}[c]{@{}c@{}} $\Sigma_{\text{SFR}}$ \\ ($M_{\odot}/\rm{yr}/\rm{kpc^{2}}$) \end{tabular}  & \begin{tabular}[c]{@{}c@{}}$\Sigma_{\text{star}}$\\ ($M_{\odot}/\rm{pc^{2}}$) \end{tabular}  & \begin{tabular}[c]{@{}c@{}}$\Sigma_{\text{atomic}}$\\ ($M_{\odot}/\rm{pc^{2}}$)\end{tabular}  \\ \hline
 218.4742 & 1.4855  & $2.58^{+0.03}_{-0.44}\times 10^{-5}$ & $6.78^{+0.08}_{-1.08}\times 10^{-1}$ & $2.40^{+0.03}_{-0.41}$  \\ 
 218.4735 & 1.4880  & $6.19^{+0.07}_{-1.05}\times 10^{-5}$ & $7.45^{+0.09}_{-1.18}\times 10^{-1}$ & $3.24^{+0.04}_{-0.55}$  \\ 
 218.4733 & 1.4828  & $5.59^{+0.06}_{-0.95}\times 10^{-5}$ & $1.17^{+0.01}_{-0.19}$ & $2.56^{+0.03}_{-0.43}$  \\ 
 218.4708 & 1.4881  & $5.14^{+0.06}_{-0.87}\times 10^{-5}$ & $7.07^{+0.08}_{-1.12}\times 10^{-1}$ & $3.06^{+0.04}_{-0.52}$  \\ 
 218.4679 & 1.4826 & $4.39^{+0.05}_{-0.75} \times 10^{-5}$ & $3.67^{+0.04}_{-0.58} \times 10^{-1}$ & $2.02^{+0.02}_{-0.34}$  \\ 
 218.4714 & 1.4855  & $5.88^{+0.07}_{-0.10}\times 10^{-5}$ & $1.18^{+0.01}_{-0.19}$ & $2.24^{+0.03}_{-0.38}$  \\ 
 218.4702 & 1.4908  & $6.63^{+0.08}_{-1.13}\times 10^{-5}$ & $4.76^{+0.05}_{-0.76}\times 10^{-1}$ & $4.19^{+0.05}_{-0.71}$  \\ 
 218.4718 & 1.4931  & $1.57^{+0.02}_{-0.27} \times 10^{-4}$ & $3.16^{+0.04}_{-0.51}\times 10^{-1}$ & $4.00^{+0.05}_{-0.68}$  \\ 
 218.4714 & 1.4751  & $2.30^{+0.03}_{-0.39}\times 10^{-5}$ & $2.42^{+0.03}_{-0.38}\times 10^{-1}$ & $2.77^{+0.03}_{-0.47}$  \\ 
 218.4705 & 1.4828  & $4.34^{+0.05}_{-0.74}\times 10^{-5}$ & $9.54^{+0.11}_{-1.51}\times 10^{-1}$ & $1.84^{+0.02}_{-0.31}$ \\ 
218.4729 & 1.4906  & $1.14^{+0.01}_{-0.19}\times 10^{-4}$ & $4.61^{+0.05}_{-0.73}\times 10^{-1}$ & $4.22^{+0.05}_{-0.72}$  \\ 
 218.4719 & 1.4957  & $2.66^{+0.03}_{-0.45}\times 10^{-5}$ & $1.13^{+0.01}_{-0.18}\times 10^{-1}$ & $2.78^{+0.03}_{-0.47}$  \\ 
218.4705 & 1.4801  & $2.93^{+0.03}_{-0.50}\times 10^{-5}$ & $6.44^{+0.07}_{-1.02}\times 10^{-1}$ & $2.71^{+0.03}_{-0.46}$  \\ 
 218.4732 & 1.4802  & $4.69^{+0.05}_{-0.80}\times 10^{-5}$ & $6.14^{+0.07}_{-0.97}\times 10^{-1}$ & $3.03^{+0.03}_{-0.52}$  \\ 
 218.4721 & 1.4777  & $4.35^{+0.05}_{-0.74}\times 10^{-5}$ & $1.28^{+0.01}_{-0.20}$ & $3.29^{+0.04}_{-0.56}$  \\

  \hline
\\
\end{tabular}
\begin{flushleft}
\textit{Notes}: Same as the Table \ref{tab:properties}.\end{flushleft}
\label{tab:properties_resolved}
\end{table}

\section{Data}

\begin{figure}
    \centering
    \includegraphics[width=0.95\columnwidth]{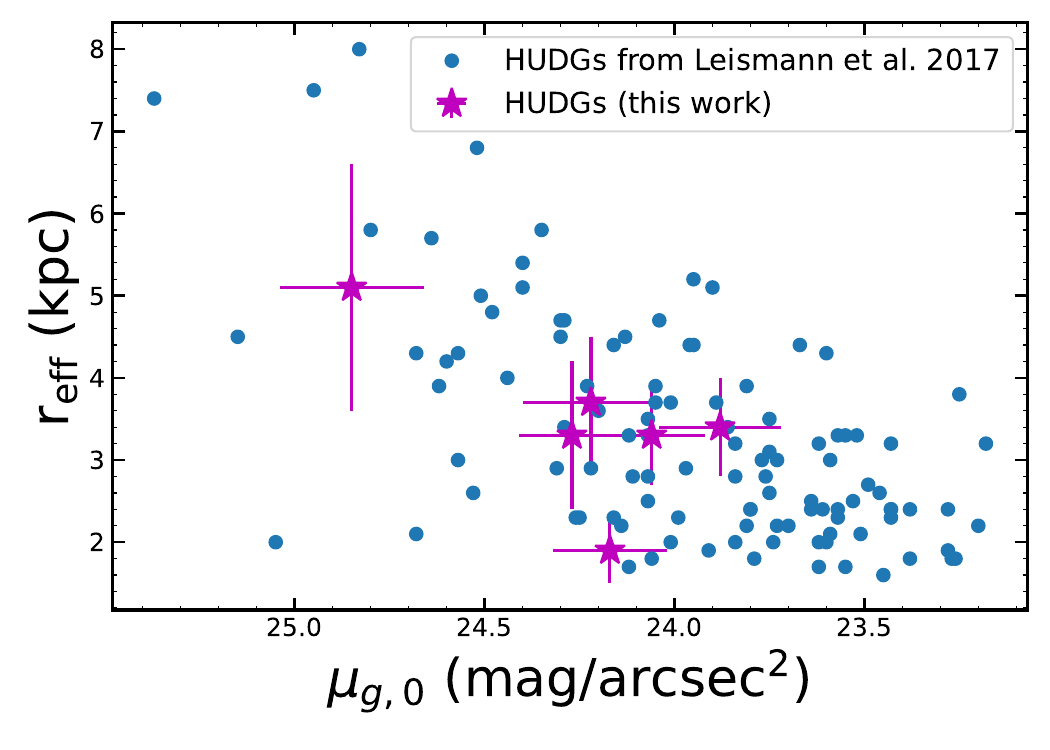}
    \caption{Optical effective radius as a function of central g band surface density. The blue dots illustrate the data of the broad criteria HUDGs from \citep{2017ApJ...842..133L}. The magenta dots represent the data of HUDGs.}
    \label{fig:ur}
\end{figure}

\begin{figure*}
    \centering
    \includegraphics[width=\textwidth]{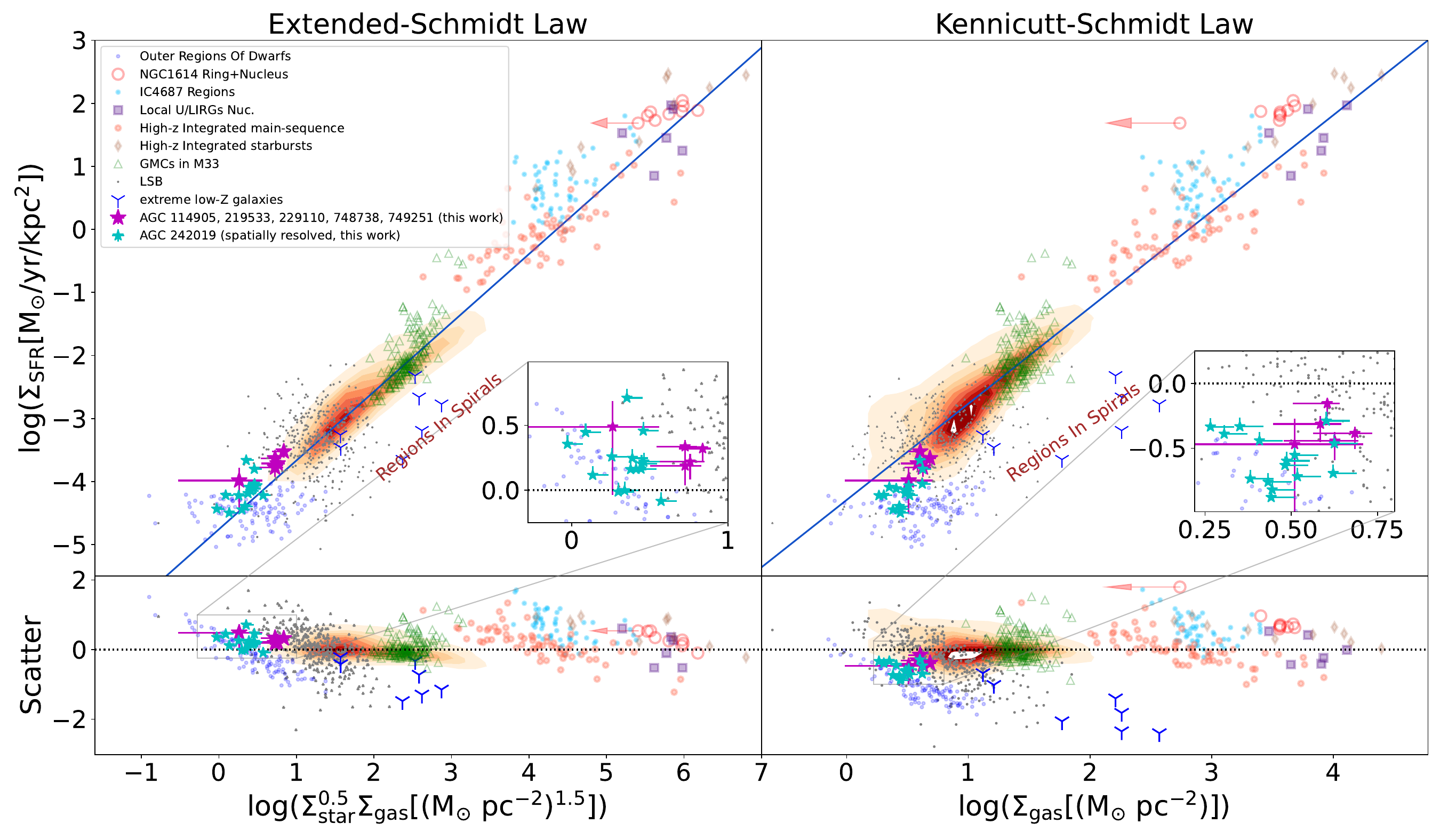}
    \caption{The extended Schmidt law: $\rm log(\Sigma_{SFR})$ as a function of $\rm log(\Sigma_{star}^{0.5}\Sigma_{gas})$. The Kennicutt-Schmidt law: $\rm log(\Sigma_{SFR})$ as a function of $\rm log(\Sigma_{gas})$. $\rm \Sigma_{gas}=\Sigma_{atomic}$ due to the undetected molecular gas in HUDGs \citep{2020MNRAS.499L..26W}. The bottom panels represent the vertical offset of the best-fitting of each star formation law. The gray line selects the magnified areas which are shown as the zoom-in plot.
    The magenta and cyan star points represent the HUDGs from \citet{2021ApJ...909...19G} and spatially resolved regions of AGC 242019, respectively. The background data points are from \citet{2014Natur.514..335S}, \citet{2018ApJ...853..149S}, \citet{2018ApJS..235...18L}, \citet{2021ApJ...909...20S} and \citet{2023MNRAS.518.4024D}. The blue solid line in the left panel and the right panel show the best fitting of the extended Schmidt law and Kennicutt-Schmidt law from \citet{2018ApJ...853..149S}, respectively.}
    \label{fig:ks}

\end{figure*}
\begin{figure}
    \centering                 
    \includegraphics[width=\columnwidth]{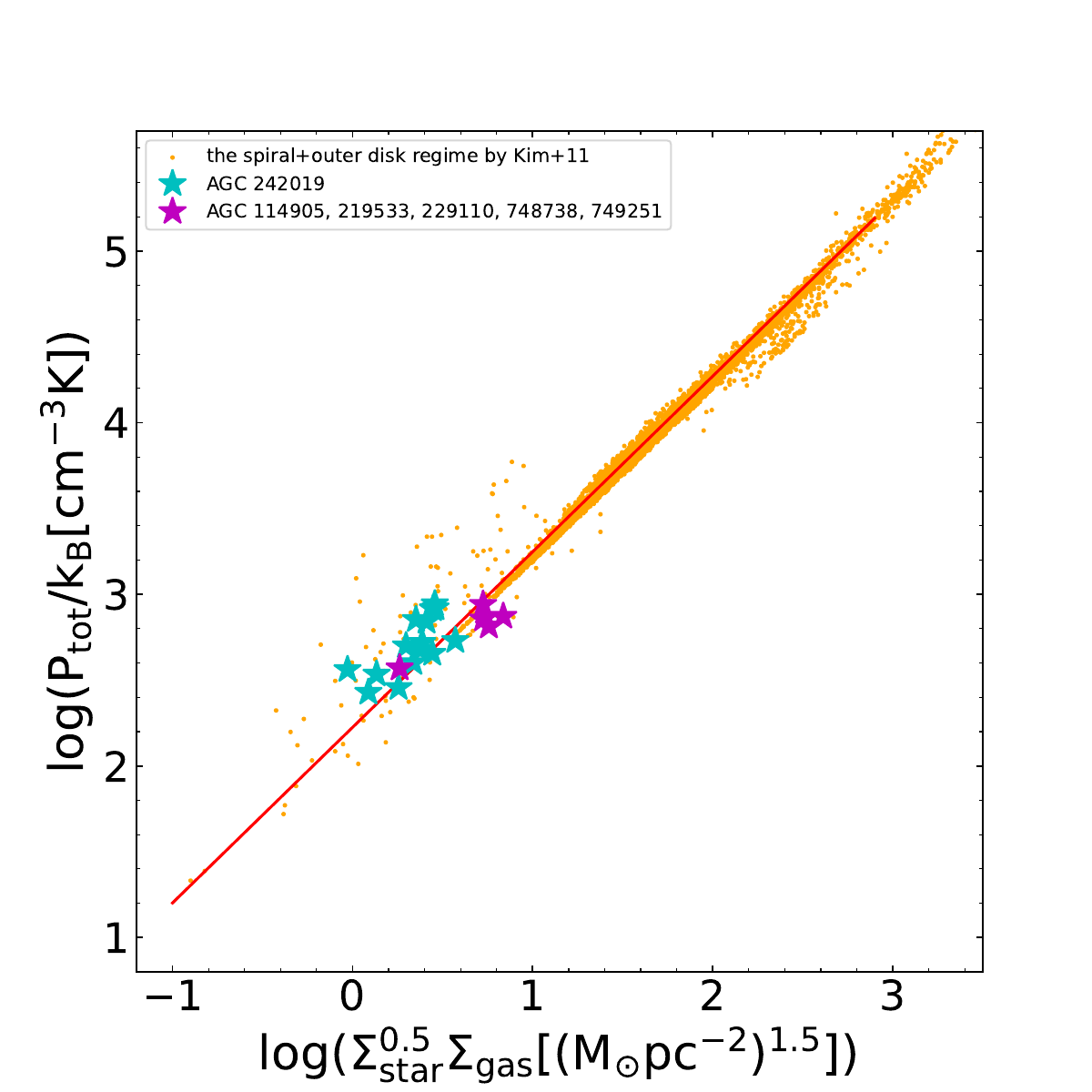} 
    \caption{The correlation between the mid-plane pressure and the $\rm log(\Sigma_{star}^{0.5}\Sigma_{gas})$.  The magenta and cyan star points represent the galaxy-wide-average HUDGs and spatially resolved regions of AGC 242019, respectively.}
    \label{fig:midplane}
\end{figure}

To select our samples of HUDGs, we require that they should have FUV band observations from the Galaxy Evolution Explorer (GALEX; \cite{2005ApJ...619L...1M}), $g$ and $r$ band images from the Dark Energy Camera Legacy Surveys \citep{2019AJ....157..168D}, and high quality spatially resolved \ion{H}{I} observations from the Karl G. Jansky Very Large Array (VLA; proposal
17A-210 \& 14B-243 (P.I. Leisman); 19B-072 \& 20A-004 (PI: Y. Shi); \cite{2017ApJ...842..133L,2021ApJ...909...19G,2021ApJ...909...20S}). Then we collect a complete sample of six HUDGs from previous work: one (AGC 242019) is from  \cite{2021ApJ...909...20S} and the others are (AGC 229110, AGC 114905, AGC 219533, AGC 748738, AGC 749251) from \cite{2021ApJ...909...19G}.

The multi-band images are shown in Fig. \ref{fig:multiband}. The photometric aperture size is predominantly defined in the g band, with the consideration of the FUV emission. Furthermore, the aperture sizes are set to be larger than the \ion{H}{I} beam size to avoid flux loss and is well larger than the point spread function (PSF) size in the other bands. For each galaxy, we use the same aperture for flux measurements in all bands. For AGC 242019, we carry out spatially resolved measurements due to its proximity along with high-quality data. The inclination angle of AGC 242019 is obtained from fitting titled-ring modelling method \citep{2015MNRAS.451.3021D,2021ApJ...909...20S}. For other HUDGs, the inclination is estimated by measuring the ratio of the minor to the major axis with DS9 \citep{2003ASPC..295..489J} based on the aperture size ($ 
\mathrm{cos(}\textit{i}\mathrm{)=b/a}$, $i$ is the inclination angle). The global and spatially-resolved properties of these galaxies are listed in Table \ref{tab:properties} and Table \ref{tab:properties_resolved}, respectively.

Following the equations (1), (2), and (3) from \cite{2018ApJ...853..149S}, we calculate the surface densities of SFR, atomic gas mass, and stellar mass, respectively. With the Kroupa initial mass function (IMF, \cite{2001MNRAS.322..231K}), we use the FUV emission from GALEX which traces the young stellar population \citep{2009ApJ...695..765M} to calculate the SFR surface density: 
\begin{equation}
\Sigma_{\text {SFR }}=8.1 \times 10^{-2} \cos (i) I_{\mathrm{FUV}},
\end{equation}
where $i$ is the inclination angle, $I_{\mathrm{FUV}}$ is the intensity of FUV band in $\rm MJy/sr$ and $\Sigma_{\text {SFR }}$ in $M_{\rm \odot} \rm  /yr/kpc^{2}$ \citep{2018ApJ...853..149S}. The dust extinction correction is not included because dwarf galaxies contain little dust in the local universe \citep{2011ApJ...741..124H}. But the extinction from the Milky Way is taken into consideration \footnote{$\rm E(B-V)$ are from \url{https://ned.ipac.caltech.edu/}} by multiplying a factor $\rm 10^{3.16E_{B-V}}$ \citep{2019ApJ...877..116W}. For $\rm \Sigma_{gas}$, the HUDGs contain little molecular gas \citep{2020MNRAS.499L..26W}, thus we approximate the \begin{equation}
\begin{split}
\rm \Sigma_{gas} \approx \rm \Sigma_{atomic} & =  1.36 \times \Sigma_{\ion{H}{I}}\\ &= 
 1.20 \times 10^{4} \cos (i)(1+z)^{3} \frac{\operatorname{arcsec}^{2}}{\mathrm{bmaj} \times \mathrm{bmin}} S_{\ion{H}{I}} \Delta v,
\end{split}
\end{equation}
where multiplying a factor of 1.36 to take the Helium into account, $\rm \Sigma_{gas}$ is the gas surface density in $M_{\rm \odot}/\rm pc^{2}$, $i$ is the inclination angle, $z$ is the redshift, $\mathrm{bmaj}, \mathrm{bmin}$ are major and minor beam size in arcsec, and  $S_{\ion{H}{I}} \Delta v$ is in $\rm Jy \ km/s/beam$ \citep{2018ApJ...853..149S}. Thus we only use the $\rm \Sigma_{atomic}$, with Helium correlation included (1.36), to represent the $\rm \Sigma_{gas}$. We use the $g$ and $r$ band photometry to calculate stellar mass ($M_{\rm star}$) following the same method as \cite{2018ApJ...853..149S}: 
\begin{equation}
\log M_{\text {star }}+M_{r-\text{band}} / 2.5=1.13+1.49(g-r),
\end{equation}
where $M_{\text {star}}$, $M_{r-\text{band}}$, and $g-r$ are the stellar mass in $M_{\odot}$, $r$ band absolute magnitude, and optical color, respectively. With the size of the aperture (A in $\rm pc^{2}$), we can obtain the stellar mass surface density:  
\begin{equation}
\rm \Sigma_{\text{star }}= M_{\text{star}}/A.
\end{equation}

\section{Results}

As shown in Fig. \ref{fig:ur}, these galaxies span a wide region in the central $g$ band surface densities ($\rm \mu_{g,0}$) and effective radius relationship ($\rm r_{eff}$)\footnote{The central $g$ band surface densities ($\rm \mu_{g,0}$) and effective radius are from \cite{2017ApJ...842..133L}. The central $g$ band surface densities ($\rm \mu_{g,0}$) are from exponential fitting. The effective radius is 1.68 times of galaxy disk scale length which contains the half light of the galaxy.}, as compared to a large sample from \cite{2017ApJ...842..133L}, which indicates that our sample in spite of the small should be representative of HUDGs.

The top left panel of Fig. \ref{fig:ks} shows that these HUDGs follow the extended Schmidt law quite well ($\Sigma_{\text {SFR }} \propto(\Sigma_{\text {star }}^{0.5} \Sigma_{\text {gas }})^{1.09}$; within the scatter). They occupy a region of the plot that is close to the outskirts of dwarf galaxies. This is found both in the disk-averaged and local measurements of HUDGs. By combining the definition of SFEs ($\text{SFE}=\Sigma_{\text{SFR}}/\Sigma_{\text{gas}}$) and the extended Schmidt law, we infer that the SFE depends on the square-root of the stellar mass surface density. For comparison, we also check the locations of these HUDGs in the Kennicutt-Schmidt law \citep{1959ApJ...129..243S,1989ApJ...344..685K} as shown in the top right panel of Fig. \ref{fig:ks}. Those HUDGs lie below the relationship. Their median vertical offset from the Kennicutt-Schmidt law is 0.47 dex (corresponding perpendicular offset: 0.32 dex), as shown in the bottom right panel of Fig. \ref{fig:ks}, which is larger than that from the extended Schmidt law (0.22 dex, corresponding perpendicular offset: 0.15 dex). This result imply that stellar mass play an important role in reducing the offset from the SFL. As a reference, the scatters of the Kennicutt-Schmidt law and the extended Schmidt law are 0.33 (corresponding perpendicular offset: 0.22 dex) and 0.22 dex (corresponding perpendicular offset: 0.15 dex), respectively \citep{2023MNRAS.518.4024D}. 
We also found that these HUDGs are located within the 1.1 sigma of ES law and 1.4 sigma of KS law. The improvement is that the ES law has a smaller sigma, while they deviate by a smaller number even in terms of sigma. We don't discussion their location at the SE relation due to the lack of rotation curve.

To address concerns regarding the sensitivity of our results to the inclination angle and aperture size, we conducted additional tests. We systematically varied the inclination angle by $\pm$ 20 degrees and adjusted the aperture size by a factor of 1.5 in the integrated data. It is found that these changes have minimal impact on the positions of the galaxies on the ES law. We observed a shift of only 0.07 dex (corresponding perpendicular offset: 0.05 dex) if altering the inclination angle and a shift of 0.04 dex (corresponding perpendicular offset: 0.03 dex) if modifying the aperture size. These findings provide evidence that our results are robust and not significantly influenced by changes in the inclination angle or aperture size. We also re-measure the aperture based on FUV emission to double-check the result. For AGC 114905 and
AGC 219533 that have reliable FUV spatially-resolved detection and their location on the ES law does not change significantly (shifted 0.04 dex, corresponding perpendicular offset: 0.03 dex). This proves the
validity of optically defined apertures.

\section{Discussion}

We discuss now the role of stellar mass in regulating star formation through the feedback self-regulated model  \citep{2010ApJ...721..975O,2009A&A...504..883B,Kim_2011,2018ApJ...853..149S,2022ApJ...936..137O,2019ApJ...871...17U,2020A&A...641A..70B}. In the star forming region of the self-regulated system, the turbulence triggered by the stellar feedback provides the pressure, which balances the gravity of the dark matter, stellar, and gas components \citep{Kim_2011,2010ApJ...721..975O}. Specifically, both the thermal pressure ($\rm P_{th}$) and turbulence pressure ($\rm P_{turb}$) contribute to the total pressure:
\begin{equation}
    \rm P_{tot}=P_{th}+P_{turb}.
\end{equation}  For self-regulating systems with gas surface densities $\rm \Sigma_{gas} = 3-20 \  M_{\odot}/\text{pc}^{2}$ and star-plus-dark matter volume densities $\rm \rho_{\text{sd}}=0.003-0.5 \  M_{\odot}/\text{pc}^{3}$, each of them is expected to be proportional to the $\rm \Sigma_{SFR}$ \citep{Kim_2011}. Because indeed the stellar components provide the majority of the feedback \citep{Kim_2011}, thus: 
\begin{equation}
    \rm P_{turb} \propto \Sigma_{SFR}.
\end{equation} And gas heating is mainly due to the radiation from massive stars \citep{Kim_2011}, implying:
\begin{equation}
    \rm P_{th} \propto \Sigma_{SFR}.
\end{equation}There should be a tight correlation between the mid-plane pressure:
\begin{equation}
    \rm P_{tot}=P_{th}+P_{turb}\propto \Sigma_{SFR} \propto (\Sigma_{star}^{0.5}\Sigma_{gas})^{1.09}. 
\end{equation} As illustrated bythe background orange dots in Fig. \ref{fig:midplane}, this relationship was confirmed to hold in spiral galaxies and outer disks of dwarfs \citep{2018ApJ...853..149S}.  The red line is the best fitting result. Following \cite{2018ApJ...853..149S}, we use the equation (35) of \cite{Kim_2011} to calculate the mid-plane pressure:

\begin{equation}
\begin{aligned}
P_{\mathrm{tot}}=& 1.7 \times 10^{3} k_{\mathrm{B}} f_{\mathrm{diff}}\left(\frac{\Sigma_{\mathrm{gas}}}{10 M_{\odot} \mathrm{pc}^{-2}}\right)^{2} \\
& \times\left\{\left(2-f_{\mathrm{diff}}\right)+\left[\left(\left(2-f_{\mathrm{diff}}\right)^{2}+37\left(\frac{\sigma_{\mathrm{z}, \mathrm{diff}}}{7 \mathrm{~km} \mathrm{~s}^{-1}}\right)^{2}\right.\right.\right.\\
&\left.\left.\times\left(\frac{\rho_{\mathrm{sd}}}{0.1 M_{\odot} \mathrm{pc}^{-3}}\right)\left(\frac{\Sigma_{\mathrm{gas}}}{10 M_{\odot} \mathrm{pc}^{-2}}\right)^{-2}\right]^{1 / 2}\right\},
\end{aligned}
\end{equation}where $k_{\mathrm{B}}$ is the Boltzmann constant that is defined to be exactly $\rm 1.380649\times10^{-23} J K^{-1}$, the mass function of diffuse components ($f_{\text {diff}}$) is fixed at $0.8$ and expressed in $\mathrm{cm}^{-3}$, and the vertical velocity dispersion in the diffuse gas ($\sigma_{z, \text { diff }}$) is fixed at $7.0$ and expressed in $\mathrm{~km} \mathrm{s}^{-1}$, and the mid-plane density of the stellar disk and dark matter (DM) halo ( $\rho_{\mathrm{sd}}$) is expressed in $M_{\odot} \mathrm{pc}^{-3}$ and further defined in \cite{2010ApJ...721..975O}:

\begin{equation}
    \rho_{\mathrm{sd}}= \rho_{\mathrm{s}} + \rho_{\mathrm{d}}= \Sigma_{\text {star }} / (2h) + \rho_{\mathrm{d}},
\end{equation}
where we assume the mid-plane density of the stellar disk $\rho_{\mathrm{s}}=\Sigma_{\text {star }} / (2h)$, and $h$ is the scale height, fixed at $1000 \mathrm{pc}$ for HUDGs. For AGC 242019, the high quality of data make it possible to derive the radial profile of the dark matter $\rho_{\text d}$ \citep{2021ApJ...909...20S}. For other HUDGs, we assume $\rho_{\mathrm{d}} \approx 0$ \citep{2010ApJ...721..975O,2022MNRAS.512.3230M}.

Then we check the behavior of HUDGs in this relationship. Figure \ref{fig:midplane} illustrates the correlation between the mid-plane pressure and $\rm log(\Sigma_{star}^{0.5}\Sigma_{gas})$. These HUDGs follow the relationship (median vertical offset: 0.14 dex, corresponding perpendicular offset: 0.10 dex), implying that they are compatible with being self-regulated systems. In this scenario, star formation is regulated by the gravitational force that compresses the gas disc along the vertical direction, influencing the gas distribution and density \citep{1985ApJ...295L...5D,2020A&A...644A.125B,2020ApJ...892..148S}. Therefore, the low SFE in HUDGs is likely due to their low stellar mass surface density.

To address the potential effect of gas mass surface density on the midplane pressure, we systematically shifted the $\rm \Sigma_{gas}$ by a factor of 1.5. These changes resulted in 0.03 dex (corresponding perpendicular offset: 0.02 dex) shift on the position of the galaxies on the Fig. \ref{fig:midplane}.

\section{Summary and Conclusions}
In this study, we collect six field HUDGs from the literature with the multiwavelength data including $g,r$, FUV, and \ion{H}{I} bands to study the origin of their low SFEs through the extended Schmidt law. 

\begin{enumerate}
\item We find that they follow the extended Schmidt law well both globally and locally, which implies the vital role of stellar mass in regulating the SFEs. 
\item We find that they follow the mid-plane pressure and $\rm log(\Sigma_{star}^{0.5}\Sigma_{gas})$ relationship, which means that the stellar mass gravity regulates the star forming process in these galaxies through mid-plane pressures.
\end{enumerate}

\section*{Acknowledgements}
S.Z. and Y.S. acknowledge the support from the National Key R\&D Program of China (No.2022YFF0503401, No. 2018YFA0404502), the National Natural Science Foundation of China (NSFC grants 11825302, 12141301, 12121003), the science research grants from the China Manned Space Project with NO. CMS-CSST-2021-B02. Y.S. thanks the Tencent Foundation through the XPLORER PRIZE. We thank the data from the proposal
17A-210 \& 14B-243 (P.I. Leisman); 19B-072 \& 20A-004 (PI: Y. Shi). We thank Lexi Gault for sharing the VLA data. We thank the  referee for constructive and detailed comments that  help improve the
paper significantly.

\section*{Data Availability}
This paper makes use of the data from VLA, GALEX, and DESI Legacy Surveys, which are available at \url{https://scholar.valpo.edu/phys_astro_fac_pub/186/}, \url{https://mast.stsci.edu/portal/Mashup/Clients/Mast/Portal.html}, and \url{https://www.legacysurvey.org/viewer/} respectively. The  \ion{H}{I} data of AGC 242019 will be shared on reasonable request to the corresponding author.



\bibliographystyle{mnras}
\bibliography{extended_ks_law} 


\bsp	
\label{lastpage}
\end{document}